\documentclass[12pt, aps,nofootinbib,superscriptaddress, showpacs,preprintnumbers,  nofootinbibt,referee]{revtex4-2}
  
\usepackage[colorlinks,linkcolor=blue,citecolor=red]{hyperref}
\usepackage{graphicx}
\usepackage{hyperref}
\usepackage{subfigure}
\usepackage{graphicx}
\usepackage{dcolumn}
\usepackage{bm}
\usepackage{amsmath,amsthm,amsfonts}
\usepackage{graphicx}
\usepackage{physics}
\usepackage{hyperref}
\usepackage{mathtools}
\usepackage{appendix}
\usepackage{comment}
\graphicspath{{figs/}}

\usepackage{amsmath}
\usepackage{subfigure}
\usepackage{natbib}
\usepackage{ulem}

\newcommand{\nc}{\newcommand}
\nc{\ba}{\begin{eqnarray}}
\nc{\ea}{\end{eqnarray}}
\nc{\bfk}{\bf{k} }
\nc{\bfq}{\bf{q} }
\nc{\rc}{\textcolor[rgb]{1.00,0.00,0.00}}
\nc{\bc}{\textcolor[rgb]{0.00,0.07,1.00}}
\newcommand{\be}{\begin{eqnarray}}
\newcommand{\ee}{\end{eqnarray}}

\def\be{\begin{equation}}
\def\ee{\end{equation}}
\def\beg{\begin{align}}
\def\eeg{\end{align}}
\def\bea{\begin{eqnarray}}
\def\eea{\end{eqnarray}}

\hypersetup{linkcolor=blue,citecolor=blue,filecolor=cyan,urlcolor=magenta}
\begin{document}

\title{Kerr black hole in presence of force-free magnetic field}

\author{Haidar Sheikhahmadi}
\email{h.sh.ahmadi@gmail.com}
\email{haidar.sheikhahmadi@nwu.ac.za}


\affiliation{School of Astronomy, Institute for Research in Fundamental Sciences (IPM),  P. O. Box 19395-5531, Tehran, Iran}
\affiliation{Center for Space Research, North-West University, Potchefstroom, South Africa,}
\begin{abstract}
We extend the study of force-free magnetospheres from non-rotating to rotating black holes \cite{Sheikhahmadi} and investigate the influence of a force-free magnetic field on the geometry around a Kerr black hole. Using the Newman-Penrose formalism, we explicitly construct the electromagnetic field strength tensor in the Kerr background and compute the corresponding stress-energy tensor. The resulting metric perturbation is then obtained by solving the linearised Einstein equations. In this modified geometry, we analyse key observables of thin accretion disks, including the innermost stable circular orbit (ISCO), effective potential, energy flux, temperature, and efficiency parameter. Our results demonstrate that magnetic backreaction significantly alters the spacetime near the black hole, with important consequences for accretion physics and jet-launching mechanisms such as the Blandford-Znajek process. This work underlines the essential role of magnetic fields in shaping relativistic astrophysical environments.
\\
\\
\textbf{Keywords:} Kerr Black Hole, Force-Free Magnetic Field, Accretion Disc, Blandford-Znajek Mechanism, Tetrad Formalism, Metric Perturbation\\
\textbf{PACS:} 97.60.Lf, 52.35.Bj, 52.65.Kj, 04.25.Nx, 04.70.Bw\\
\end{abstract}

\maketitle
\newpage
\tableofcontents

\section{Introduction}\label{introduction00}

Kerr black holes, the rotating solutions to the vacuum Einstein field equations, represent the fundamental theoretical model for astrophysical black holes \cite{Kerr1963, Newman1965, Boyer1966}. Observations, particularly by the Event Horizon Telescope collaboration, have provided direct evidence for the existence of such objects and their associated phenomena \cite{EHT2019M87, EHT2022SgrA}. In their extreme gravitational fields, the interplay between gravity, electromagnetism, and plasma physics leads to complex behavior, with force-free magnetospheres playing a particularly critical role 
\citep{Chandrasekhar01, ChandraPapers-RNpapereb, Chandrasekhar03, Ref16J, Ref16Ja, Ref16Jb, Jac}.

The seminal Blandford-Znajek mechanism describes how a rotating black hole can interact with a surrounding magnetic field to power relativistic jets, offering a primary explanation for the energetic outflows observed in active galactic nuclei and other systems \cite{BlandfordZnajek1977}. In this model, the magnetic field is assumed to be force-free, a valid approximation when the electromagnetic energy density dominates the plasma's inertial and thermal energies \cite{Goldreich1969, Gruzinov1999}. In this regime, the Lorentz force vanishes, and the plasma becomes perfectly conducting, with the magnetic field dictating the dynamics.

The study of force-free magnetic fields (FFMFs) in Kerr spacetime is therefore essential for understanding accretion physics, jet launching, and energy extraction \cite{McKinney2004, Tchekhovskoy2011}. Recent analytical and numerical works have advanced our understanding of these magnetospheres, exploring their stationary configurations and stability \cite{Uzdensky2004, Contopoulos2013,Gralla2017,ChandraBookHydro1,Aschwanden,Uzdensky:2004qu,Contopoulos:2012py,Lupsasca:2014hua}. However, a consistent treatment requires considering the backreaction of these electromagnetic fields on the spacetime geometry itself, moving beyond the test-field approximation on a fixed background. Here we should emphasize that besides the detailed literature mentioned above, a numerous list of rich research works is devoted to FFMFs in different backgrounds; we refer the reader to some of them
\cite{Wilson-Gerow:2015esa, Pan:2015iaa, Huang:2020lvl, Grignani:2018ntq,Zajacek:2018ycb, ChandraPapers-RNpaperg, ChandarbookBH,Yuan:2019mdb, Grignani:2019dqc, Camilloni:2020hns,Mirabel2011,Marsh,Priest}.

{ While recent investigations have} provided solutions for FFMFs around Schwarzschild black holes and analyzed their metric-modifying effects \cite{Sheikhahmadi}, the extension to the rotating Kerr case, which is astrophysically paramount, presents significant additional complexity. This work aims to bridge this gap by deriving a self-consistent perturbative solution for a Kerr black hole immersed in a force-free magnetosphere. We employ the tetrad formalism to formulate the problem covariantly and compute the electromagnetic stress-energy tensor that sources the metric perturbations. It is worth noting here that, by utilizing the Newman-Penrose formalism, one can make a connection between curved backgrounds and their tangent space vierbeins, i.e., tetrads \cite{Newman1962,Teukolsky1973}.

Our analysis demonstrates that a perturbation theory which neglects the gravitational backreaction of the magnetofluid is fundamentally incomplete. The modified spacetime geometry, in turn, influences the magnetic field structure, necessitating a coupled approach. This work thus provides a foundational step towards a more complete, non-linear understanding of magnetized Kerr black holes.

The paper is structured as follows: Section \ref{Force Free Magnetic Field} details the mathematical framework of force-free electrodynamics and the tetrad formalism in curved spacetime. Section \ref{EM-Tensor-Kerr} derives the explicit form of the electromagnetic stress-energy tensor in the Kerr background. Using this as a source, Section \ref{Perturbationshmunu} solves the linearized Einstein equations to obtain the metric perturbation. Sections \ref{ParticleMotion} and \ref{Observ00} compare the results to the normal Kerr metric to see the effects of the FFMF on geodesics, flux, thermal properties of the accretion disk, and other physical properties of such a configuration. Finally, Section \ref{Final} discusses the physical implications of our results and presents concluding remarks, with a summary of the key figures that illustrate our findings.

\section{Mathematics of force-free magnetic fields}
\label{Force Free Magnetic Field}

We investigate force-free electrodynamics (FFE) within the curved spacetime of a rotating compact object, i.e., a Kerr black hole. Our analysis begins by formulating the fundamental equations in spherical coordinates and subsequently employs the Newman-Penrose tetrad formalism \cite{Cartan1922, Newman1962} to transform the electromagnetic field strength tensor into a form compatible with the Kerr geometry. This covariant approach is essential for handling the complexities introduced by frame-dragging and spacetime curvature.

FFE provides the theoretical basis for modeling key astrophysical processes driven by magnetized black holes. Most notably, the Blandford-Znajek mechanism \cite{BlandfordZnajek1977}, operating within this framework, explains the extraction of rotational energy to power relativistic jets, such as those observed from the supermassive black hole in M87 \cite{EventHorizonTelescope:2019ggy}, and may contribute to the central engine of gamma-ray bursts \cite{Lyman:2018qjg}. While the core force-free principles are universally applicable, the specific magnetic field configuration and boundary conditions are context-dependent, shaped by the properties of the accretion flow and the global magnetospheric structure \cite{Tchekhovskoy2011, Contopoulos2013}.

The central aim of this work is to compute the gravitational backreaction of the force-free magnetosphere on the black hole spacetime. We treat the stress-energy tensor 
\begin{equation}
    T_{\mu\nu}^{\mathrm{(EM)}} = \frac{1}{4\pi}\left(F_{\mu\alpha}F_{\nu}^{\;\alpha} - \frac{1}{4}g_{\mu\nu}F_{\alpha\beta}F^{\alpha\beta}\right),
    \label{eq:em_stress_energy}
\end{equation}
of the force-free magnetofluid as a source term in the linearized Einstein field equations, thereby deriving the metric perturbations \(h_{\mu\nu}\) to the background Kerr solution. Studying this coupled system, where the modified spacetime geometry influences the magnetic field structure and vice versa, is critical for developing a self-consistent model of magnetized black holes and for accurately predicting observable signatures from accretion disks and jets.

\subsection{Force-free fields in spherical coordinates}\label{Spherical}

To establish our approach, we first derive the components of a force-free magnetic field in a flat spacetime background, formulated in spherical coordinates. The Minkowski metric is
\begin{equation}\label{flat-metric}
d{S^2} = -d{t^2} + d{r^2} + {r^2}d{\theta ^2} +{r^2}{\sin ^2}\theta\, d{\varphi ^2}\,.
\end{equation}
In the force-free approximation, the magnetic field \(\mathbf{B}\) satisfies the Helmholtz equation
\begin{equation}\label{2-1}
{\nabla ^2}\mathbf{B} + {\alpha ^2}\mathbf{B} = 0\,,
\end{equation}
where \(\alpha\) is a scalar function. For the tractable case of constant \(\alpha\), this equation admits an exact solution. Following the established approach \cite{ChandraPapers-RNpaperc, Sobouti1}, we introduce a scalar potential \(\Psi\) such that
\begin{equation}\label{2-1Psi}
{\nabla ^2}\Psi + {\alpha ^2}\Psi = 0\,.
\end{equation}
This scalar equation admits three independent vector solutions:
\begin{equation}\label{4-1Fa}
 \mathbf{L}_S = \nabla \Psi \,, \quad \mathbf{T} = \nabla  \times (\hat{e}\Psi )\,, \quad \mathbf{S} = \frac{1}{\alpha }\nabla  \times \mathbf{T}\,,
\end{equation}
known respectively as the solenoidal, toroidal, and poloidal components. Here \(\hat{e}\) is an arbitrary constant unit vector defining an axis of symmetry. A general force-free magnetic field can then be constructed as
\begin{equation}\label{4-1B}
\mathbf{B} =  \mathbf{T}+\mathbf{S} = \frac{1}{\alpha }\nabla  \times \nabla  \times (\hat{e}\Psi ) + \nabla  \times (\hat{e}\Psi )\,.
\end{equation}

Assuming separability, \(\Psi(r, \theta)=R(r)\Theta(\theta)\), and using the flat metric (\ref{flat-metric}), Eq. (\ref{2-1Psi}) decouples into radial and angular equations:
\begin{eqnarray}\label{7-1}
\frac{1}{{{r^2}}}\frac{d}{{dr}}\left({{{r^2}}}\frac{{dR(r)}}{{dr}}\right) + \left({\alpha ^2} - \frac{{n(n + 1)}}{{{r^2}}}\right)R(r) &=0 \,, \cr
\frac{1}{{\sin \theta }}\frac{d}{{d\theta }}\left(\sin \theta \frac{{d\Theta (\theta )}}{{d\theta }}\right) + n(n + 1)\Theta (\theta ) &=0\, .
\end{eqnarray}
The general solution for \(\Psi\) is a sum over modes:
\begin{equation}\label{5-1FbK}
    \Psi(r,\theta)=\sum_{n=1}^{\infty} (k r)\, j_{n}(k r)\,(1-\cos \theta)\, P_{n}^{(1,-1)}(\cos \theta)\,,
\end{equation}
where \( j_{n}\) are spherical Bessel functions, \(P_{n}^{(1,-1)}\) are Jacobi polynomials, and
 \(k = \frac{\alpha}{\Psi}\left(K^{2}+\Psi^{2}\right)^{1 / 2}\). For a simple constant \(\alpha\) (e.g., \(K=0\)), we have \(k=\alpha\) \cite{Marsh, Sheikhahmadi}. The lowest-order mode (\(n=1\)) yields the simple expression
\begin{equation}\label{8-1}
\Psi(r, \theta)=\left( \frac{{\sin (k r)}}{{{{k r}}}} - {{\cos (k r)}}\right) \sin^2 \theta\,.
\end{equation}
Using this potential, the components of an axisymmetric magnetic field in spherical coordinates become
\begin{equation}\label{MagneticBr}
  {B_r} = \frac{1}{{{r^2}\sin \theta }}{\partial _\theta }\Psi\,, \quad {B_\theta } =  - \frac{1}{{r\sin \theta }}{\partial _r}\Psi\,, \quad {B_\phi } = \frac{k\Psi}{{r\sin \theta }}\,.
\end{equation}

In the force-free magnetosphere around a black hole, the electric field in the fluid frame vanishes. Therefore, we set \({E_r} = {E_\theta } = {E_\phi } = 0\) \cite{Blinder:2004ik}. Given the magnetic field components and the metric (\ref{flat-metric}), the covariant and contravariant forms of the electromagnetic field tensor \(F_{ab}\) are:
\begin{equation}\label{Falhpabeta}
  {\mathfrak{F}_{a b }} = \left( {\begin{array}{*{20}{c}}
0&0&0&0\\
0&0& - r{B_\phi }& r\sin \theta {B_\theta }\\
0& r{B_\phi }&0& - {r^2}\sin \theta {B_r}\\
0& - r\sin \theta {B_\theta }&{r^2}\sin \theta {B_r}&0
\end{array}} \right)\,,
\end{equation}
\begin{equation}\label{FCont-alhpabeta}
  {\mathfrak{F}^{a b }} =\left( {\begin{array}{*{20}{c}}
0&0&0&0\\
0&0& - \frac{{{B_\phi }}}{r}& \frac{{{B_\theta }}{}}{{r\sin \theta }}\\
0& \frac{{{B_\phi }}}{r}&0& - \frac{{{B_r}}}{{{r^2}\sin \theta }}\\
0& - \frac{{{B_\theta }}{}}{{r\sin \theta }}& \frac{{{B_r}}}{{{r^2}\sin \theta }}&0
\end{array}} \right)\,,
\end{equation}
where Latin indices \(a, b\) denote coordinates in the spherical basis \((t, r, \theta, \phi)\). This flat-space solution provides the foundational structure which we will now generalize to the Kerr spacetime using the tetrad formalism.

\subsection{Force-free fields in the Kerr metric}\label{Kerr-munu}

We now extend the analysis to the rotating Kerr spacetime, described in Boyer-Lindquist coordinates \cite{Boyer1966} by the line element
\begin{equation}\label{Kerr-B-L}
d{s^2} =  - \frac{\Delta }{{\rho ^2}}{\left( {dt - a\sin^2\theta\, d\phi } \right)^2} + \frac{\sin^2\theta }{{\rho ^2}}{\Big( {\left( {{r^2} + {a^2}} \right)d\phi  - a\,dt} \Big)^2} + \frac{{\rho ^2}}{\Delta }d{r^2} + {\rho ^2}d{\theta ^2}\,,
\end{equation}
where
\begin{equation}
\Delta  = {r^2} - 2Mr + {a^2}, \qquad \rho^2 = {r^2} + {a^2}\cos^2\theta\,,
\end{equation}
with \(a = J/M\) being the Kerr spin parameter, \(M\) the mass, and we set \(G = c = 1\). For clarity in subsequent tensor calculations, we present both the covariant and contravariant forms of the metric. The indices \(\mu, \nu = 1, 2, 3, 4\) correspond to the coordinates \((r, \theta, \phi, t)\), where the fourth index denotes the time component. In the matrices below, the \textbf{boldface} components explicitly indicate the time-time elements \(g_{44}\) and \(g^{44}\). The covariant metric tensor is
\begin{equation}\label{Kerr-main}
{g_{\mu \nu }} = \left( {\begin{array}{*{20}{c}}
\frac{\rho^2}{\Delta}&0&0&0\\
0&\rho^2&0&0\\
0&0&\frac{{{{\sin }^2}(\theta )\left( {{r^2}{\rho ^2} + {a^2}(\Delta  - {a^2}{{\sin }^2}(\theta ) + 2Mr{{\cos }^2}(\theta ))} \right)}}{{{\rho ^2}}}&{ - \frac{{2Mra{{\sin }^2}(\theta )}}{\rho^2}}\\
0&0&{ - \frac{{2Mra{{\sin }^2}(\theta )}}{\rho^2}}&{-\frac{{{\bf{\Delta}} - {{\bf{a}}^{\bf{2}}}{{\sin }^{\bf{2}}}(\theta )}}{\mathbf{\rho^2}}}
\end{array}} \right)\,.
\end{equation}
The contravariant form of the metric (\ref{Kerr-main}) then reads
\begin{equation}\label{Kerr-Inverse}
 g^{\mu\nu}= \left( {\begin{array}{*{20}{c}}
\frac{\Delta}{\rho^2}&0&0&0\\
0&{\frac{1}{\rho^2}}&0&0\\
0&0&\frac{{\left( {\Delta  - {a^2}{{\sin }^2}(\theta )} \right){{\csc }^2}(\theta )}}{{\Delta {\rho ^2}}}& - \frac{{2aMr}}{{\Delta {\rho ^2}}}\\
0&0& - \frac{{2aMr}}{{\Delta {\rho ^2}}}&\mathbf{ - \frac{{{r^2}{\rho ^2} + {a^2}(\Delta  - {a^2}{{\sin }^2}(\theta ) + 2Mr{{\cos }^2}(\theta ))}}{{\Delta {\rho ^2}}}}
\end{array}} \right)\,.\\
\end{equation}

To transform the electromagnetic field tensor \(\mathfrak{F}_{ab}\) from the local spherical basis (Sec.~\ref{Spherical}) to the Kerr background, we employ the tetrad formalism \cite{ChandarbookBH, Newman1965}. The transformation is effected via the tetrads \(e^a_\mu\), which satisfy \(g_{\mu\nu} = e^a_\mu e^b_\nu \, \eta_{ab}\), where \(\eta_{ab}\) is the local Minkowski metric. The electromagnetic tensor in the curved spacetime is then
\begin{equation}\label{F to F}
 {F_{\mu \nu }} = e_\mu ^a e_\nu ^b {\mathfrak{F}_{ab}}\,.
\end{equation}
A set of tetrads suitable for this computation is
\begin{equation}\label{Tetrads}
\begin{aligned}
e_{\mu=1}^{a=1} &= \frac{\rho}{\sqrt{\Delta}}, \quad &
e_{\mu=2}^{a=2} &= \rho, \\
e_{\mu=3}^{a=3} &= \frac{\rho \sin\theta \sqrt{\Delta}}{\sqrt{a^2\cos^2\theta - 2Mr + r^2}}, \quad &
e_{\mu=4}^{a=4} &= \frac{\sqrt{a^2\cos^2\theta - 2Mr + r^2}}{\rho}\,.
\end{aligned}
\end{equation}

Using the flat-space field tensor components from Eq.~(\ref{Falhpabeta}) with the tetrads in Eq.~(\ref{Tetrads}) and the transformation rule (\ref{F to F}), we compute the non-zero covariant components of \(F_{\mu\nu}\) in the Kerr geometry. For the magnetic field derived from the potential \(\Psi\) in Eq.~(\ref{8-1}), these are:
\begin{equation}\label{F12-Kerr}
{F_{12}} = \frac{{\rho^2 \sin \theta \left( {\cos \left( {k r} \right) k r - \sin \left( {k r} \right)} \right)}}{{\sqrt{\Delta} \, r}}\,,
\end{equation}
\begin{equation}\label{F13-Kerr}
{F_{13}} = - \frac{{\rho^2 \sin^2 \theta \left( {{k^2}\sin \left( {k r} \right) r^2 + \cos \left( {k r} \right) k r - \sin \left( {k r} \right)} \right)}}{{\sqrt{a^2\cos^2\theta - 2Mr + r^2} \, k r^2}}\,,
\end{equation}
\begin{equation}\label{F23-Kerr}
{F_{23}} = \frac{{2 \rho^2 \sin \theta \sqrt{\Delta} \left( {\cos \left( {k r} \right) k r - \sin \left( {k r} \right)} \right) \cos \theta }}{{\sqrt{a^2\cos^2\theta - 2Mr + r^2} \, k r}}\,.
\end{equation}
The corresponding contravariant components are obtained via the tensor transformation rule \cite{Jackson}
\begin{equation}\label{F-contra}
{F^{\alpha \beta }} = {g^{\alpha \mu }}{g^{\beta \nu }}{F_{\mu \nu }}\,,
\end{equation}
yielding
\begin{equation}\label{F12conta-Kerr}
{F^{12}} = - \frac{{\sqrt{\Delta} \sin \theta \left( {\cos \left( {k r} \right) k r - \sin \left( {k r} \right)} \right)}}{{\rho^2 \, r}}\,,
\end{equation}
\begin{equation}\label{F13conta-Kerr}
{F^{13}} = - \frac{{\sqrt{a^2\cos^2\theta - 2Mr + r^2} \left( {{k^2}\sin \left( {k r} \right) r^2 + \cos \left( {k r} \right) k r - \sin \left( {k r} \right)} \right)}}{{\rho^2 \, k r^2}}\,,
\end{equation}
\begin{equation}\label{F23conta-Kerr}
{F^{23}} = - \frac{{2 \sqrt{a^2\cos^2\theta - 2Mr + r^2} \left( { - \cos \left( {k r} \right) k r + \sin \left( {k r} \right)} \right) \cot \theta }}{{\rho^2 \sqrt{\Delta} \, k r}}\,.
\end{equation}
With the electromagnetic field tensor fully determined in the Kerr background, we can now compute its stress-energy tensor, \(T_{\mu\nu}^{\text{(EM)}}\), which will serve as the source term for perturbing the background metric in the following section.

\section{FFMF energy momentum tensor as a source of perturbations}\label{EM-Tensor-Kerr}

First we will obtain components of the energy momentum tensor, then we will solve the perturbed Einstein field equations to obtain the modifications in the Kerr metric induced by the FFMF source.

\subsection{Energy momentum tensor}\label{EM-Dia-Off-Dia}

The definition of the energy momentum tensor in a general curved background reads \citep{Jackson}
\begin{equation}\label{E-M01}
   {{T}_{\mu \nu }} = {F_{\mu \gamma }}F_\nu ^\gamma  - \frac{1}{4}g_{\mu\nu}^{(0)}({F_{\alpha\beta}}{F^{\alpha\beta }})\,,
\end{equation}
where \(g_{\mu\nu}^{(0)}\) here is the background Kerr metric, Eq.(\ref{Kerr-main}). The diagonal components, for our problem, read
\begin{equation}\label{Em-Compos}
\begin{array}{l}
{T_{44}} = \frac{{ - {g^{(0)}_{44}}}}{2}({F_{12}}{F^{12}} + {F_{13}}{F^{13}} + {F_{23}}{F^{23}});\,\,\,\,{T_{11}} = \frac{{{g^{(0)}_{11}}}}{2}({F_{12}}{F^{12}} + {F_{13}}{F^{13}} - {F_{23}}{F^{23}})\,,\\
\\
{T_{22}} = \frac{{{g^{(0)}_{22}}}}{2}({F_{12}}{F^{12}} - {F_{13}}{F^{13}} + {F_{23}}{F^{23}});\,\,\,\,{T_{33}} = \frac{{{g^{(0)}_{33}}}}{2}( - {F_{12}}{F^{12}} + {F_{13}}{F^{13}} + {F_{23}}{F^{23}})\,.
\end{array}
\end{equation}
For the off-diagonal components we obtain
\begin{equation}\label{Off-Diag}
\begin{array}{l}
{T_{43}} = {T_{34}} =  - \frac{{{g^{(0)}_{34}}}}{2}({F_{12}}{F^{12}} + {F_{13}}{F^{13}} + {F_{23}}{F^{23}})\\
{T_{12}} = {g^{(0)}_{22}}{F_{13}}{F^{23}};\,\,{T_{21}} = {g^{(0)}_{11}}{F_{23}}{F^{13}}\,,\\
{T_{13}} = {g^{(0)}_{33}}{F_{12}}{F^{32}};\,\,{T_{31}} = {g^{(0)}_{11}}{F_{32}}{F^{12}}\,,\\
{T_{23}} = {g^{(0)}_{33}}{F_{21}}{F^{31}};\,\,{T_{32}} = {g^{(0)}_{22}}{F_{31}}{F^{21}}\,.
\end{array}
\end{equation}

By using the results of Eqs.(\ref{F12-Kerr}) to (\ref{F23conta-Kerr}) one gets
\begin{eqnarray}\label{T44-Kerr}\nonumber
T_{44} &= &-\frac{{\left( {\Delta  - {a^2}{{\sin }^2}(\theta )} \right){{\sin }^2}\theta }}{{2{\rho ^2}}}\Bigg(\frac{{{{\sin }^2}\theta {{\big( {\cos \left( {kr} \right)kr - \sin \left( {kr} \right)} \big)}^2}}}{{{r^2}}} \\\nonumber
&+&  \frac{{4{{\cos }^2}\theta {{\big( {\cos \left( {kr} \right)kr - \sin \left( {kr} \right)} \big)}^2}}}{{{k^2}{r^2}}}+\frac{{{{\sin }^2}\theta {{\big( {{k^2}\sin \left( {kr} \right){r^2} + \cos \left( {kr} \right)kr - \sin \left( {kr} \right)} \big)}^2}}}{{{k^2}{r^4}}}\Bigg) \,,\\
\end{eqnarray}
\begin{eqnarray}\label{T11-Kerr}\nonumber
T_{11} &=&- \frac{{\left( {\Delta  - {a^2}{{\sin }^2}(\theta )} \right){{\sin }^2}\theta }}{{4{\rho ^2}}}\Bigg(\frac{{{{\big( {\cos \left( {kr} \right)kr - \sin \left( {kr} \right)} \big)}^2}}}{{{r^2}}} \\
&-& \frac{{4{{\cot }^2}\theta {{\big( {\cos \left( {kr} \right)kr - \sin \left( {kr} \right)} \big)}^2}}}{{{k^2}{r^2}}}+ \frac{{{{\big( {{k^2}\sin \left( {kr} \right){r^2} + \cos \left( {kr} \right)kr - \sin \left( {kr} \right)} \big)}^2}}}{{{k^2}{r^4}}}  \Bigg){\mkern 1mu},
\end{eqnarray}
\begin{eqnarray}\label{T22-Kerr}\nonumber
T_{22} &=&- \frac{{\left( {\Delta  - {a^2}{{\sin }^2}(\theta )} \right){{\sin }^2}\theta }}{{4}}\Bigg(\frac{{{{\big( {\cos \left( {kr} \right)kr - \sin \left( {kr} \right)} \big)}^2}}}{{{r^2}}} \\
&-& \frac{{4{{\cot }^2}\theta {{\big( {\cos \left( {kr} \right)kr - \sin \left( {kr} \right)} \big)}^2}}}{{{k^2}{r^2}}}+ \frac{{{{\big( {{k^2}\sin \left( {kr} \right){r^2} + \cos \left( {kr} \right)kr - \sin \left( {kr} \right)} \big)}^2}}}{{{k^2}{r^4}}}  \Bigg){\mkern 1mu},
\end{eqnarray}
\begin{eqnarray}\label{T33-Kerr}\nonumber
T_{33}&=&\frac{{{{\sin }^4}\theta \left( {\frac{{2Mr{\mkern 1mu} {a^2}{{\sin }^2}\theta }}{{ \rho^2}} + {r^2} + {a^2}} \right)}}{2} \Bigg( \frac{{{{\left( {\cos \left( {kr} \right)kr - \sin \left( {kr} \right)} \right)}^2}}}{{{r^2}}}\\
& + &\frac{{{{\left( {{k^2}\sin \left( {kr} \right){r^2} + \cos \left( {kr} \right)kr - \sin \left( {kr} \right)} \right)}^2}}}{{{k^2}{r^4}}} + \frac{{4{{\cot }^2}\theta {{\left( {\cos \left( {kr} \right)kr - \sin \left( {kr} \right)} \right)}^2}}}{{{k^2}{r^2}}} \Bigg)\,.
\end{eqnarray}
Avoiding prolongation of the expressions, we bring only some of the off-diagonal components, for instance
\begin{eqnarray}\label{T43-Kerr}\nonumber
{T_{43}} = {T_{34}}&=&\frac{{M\,r{\mkern 1mu} a{{\sin }^4}\theta }}{{2\rho^2}} \Bigg( \frac{{{{\left( {\cos \left( {kr} \right)kr - \sin \left( {kr} \right)} \right)}^2}}}{{{r^2}}}\\\nonumber
& + &\frac{{{{\left( {{k^2}\sin \left( {kr} \right){r^2} + \cos \left( {kr} \right)kr - \sin \left( {kr} \right)} \right)}^2}}}{{{k^2}{r^4}}} + \frac{{4{{\cot }^2}\theta {{\left( {\cos \left( {kr} \right)kr - \sin \left( {kr} \right)} \right)}^2}}}{{{k^2}{r^2}}} \Bigg)\,.\\
\end{eqnarray}
\begin{equation}\label{T13-Kerr}
{T_{13}} = {T_{31}}={\sin^2} \theta \cos\theta\Bigg( \frac{{{{\cos }^2}\left( {kr} \right)kr}}{2} - \frac{{\cos \left( {kr} \right)\sin \left( {kr} \right)}}{2}\, + \left( {\frac{{m\cos \left( {kr} \right)}}{2} - \frac{{\sin \left( {kr} \right)}}{{2k}}} \right)k\cos \left( {kr} \right)\Bigg)\,,
\end{equation}
\begin{eqnarray}\label{T12-Kerr}\nonumber
{T_{12}}={T_{21}} & =&  - \frac{{\rho^2 \cot \theta}}{{2{k^2}{r^3}\sqrt {{a^2} - 2Mr + {r^2}} }}\\
&~&\Big( {{k^2}\sin \left( {kr} \right){r^2} + \cos \left( {kr} \right)kr - \sin \left( {kr} \right)} \Big)\big( {\cos \left( {kr} \right)kr - \sin \left( {kr} \right)} \big) \,.
\end{eqnarray}

\subsection{FFMF as a source of perturbations on Kerr background}\label{Perturbation-FFMF}
Here, to study the effects of the induced FFMF on the Kerr black hole, we employ the well-established linearized theory of gravity \citep{Weinberg,MTW,ChandraPapers-RNpaperf,Ryder}. In this approach one can perturbatively write down the general form of the metric as
\begin{equation}\label{1-0}
g_{\mu\nu}=g_{\mu\nu}^{(0)}+h_{\mu\nu},
\end{equation}
where the source of \(h_{\mu\nu}\) arises from force-free magnetohydrodynamics and \(g_{\mu\nu}^{(0)}\) refers to the vacuum Kerr solution. We must emphasize here that \(g_{\mu\nu}^{(0)}\) is used to raise and lower indices in the linearized gravity approach. Accordingly, the Einstein field equations can be expressed as follows
\begin{equation}\label{Einsteinmunu}
G_{\mu \nu }^{(0)} + \delta {G_{\mu \nu }} = 0 + {T_{\mu \nu }}\,.
\end{equation}
where
\begin{equation}\label{Einsteinmunu2}
 \delta {G_{\mu \nu }} = \delta {R_{\mu \nu }}-\frac{1}{2}g_{\mu\nu}^{(0)} \delta {R}\,.
\end{equation}
We consider the first order of perturbation; in doing so we need to calculate the first-order
Christoffel symbols, Riemann and Ricci tensors and consequently the Ricci scalar.

For the metric (\ref{1-0}), the Christoffel symbols read
 \begin{equation}\label{Christoffel 01}
\Gamma_{\mu\nu}^{(1)\eta}=\frac{1}{2}\left(\nabla_{\nu} h_{\mu}^{\eta}+\nabla_{\mu} h_{\nu}^{\eta}-\nabla^{\eta} h_{\nu \mu}\right)\,.
\end{equation}
By utilizing Eq.(\ref{Christoffel 01}), one easily gets the Riemann tensor as follows
\begin{equation} \label{Riemann01}
\begin{array}{l}
R_{\mu \nu \sigma}^{(1) \eta}=\frac{1}{2}\left(\nabla_\nu \nabla_\sigma h_\mu^\eta+\nabla_\nu \nabla_\mu h_\sigma^\eta-\nabla_\nu \nabla^\eta h_{\mu \sigma}-\nabla_\sigma \nabla_\nu h_\mu^\eta-\nabla_\sigma \nabla_\mu h_\nu^\eta+\nabla_\sigma \nabla^\eta h_{\mu \nu}\right),
\end{array}
\end{equation}
then by using \(g_{\mu\nu}^{(0)}\) to contract Eq.(\ref{Riemann01}), the Ricci tensor is obtained as
\begin{equation}\label{Ricci01}
R_{\mu\nu}^{(1)}=R_{\mu \eta \nu}^{(1) \eta}=\frac{1}{2}\left(\nabla_{\eta} \nabla_{\nu} h_{\mu}^{\eta}+\nabla_{\eta} \nabla_{\mu} h_{\nu}^{\eta}-\nabla_{\eta} \nabla^{\eta} h_{\mu k}-\nabla_{\nu} \nabla_{\mu} h\right)\,,
\end{equation}
where \(h\) is the trace of \(h_{\mu\nu}\). To obtain the Ricci scalar the following relation is constructive
\begin{equation}\label{RicciScalar01}
R_{\nu}^{(1)\mu}=g^{\mu \eta(0)} R_{\nu \eta}^{(1)}-h^{\mu \eta} R_{\nu \eta}^{(0)}\,.
\end{equation}
To simplify our analysis, we find it more convenient to use a modified metric called the trace-reversed metric, denoted as \(\bar{h}_{\mu\nu}\). This metric is related to the perturbed metric \(h_{\mu\nu}\) as follows
\begin{equation}\label{hbar01}
\bar{h}_{\mu \nu} \equiv h_{\mu \nu} - \frac{1}{2} g_{\mu \nu}^{(0)} h\,.
\end{equation}
Additionally, we introduce a small coordinate transformation represented by \(x^\mu \longrightarrow x^\mu + \xi^\mu\) \citep{ChandraPapers-RNpaperf}. By imposing the Lorentz gauge condition, we obtain an important equation that describes the propagation of an electromagnetic wave on a curved background, given by
\begin{equation}\nonumber
\square \xi^{\mu} = -\nabla_{\nu} \bar{h}^{\mu\nu}.
\end{equation}
This equation describes the relationship between the transformation \(\xi^\mu\) and the trace-reversed metric components. It provides valuable insights into how electromagnetic waves propagate in a curved spacetime. In fact, we can express it as follows
\begin{equation}\label{wavepropagation01}
\square \bar{h}_{\mu \nu} + 2 R_{\eta \mu \epsilon \nu}^{(0)} \bar{h}^{\eta \epsilon} = 4 {T}_{\mu\nu},
\end{equation}
Here, \(R_{\eta \mu \epsilon \nu}^{(0)}\) represents the Riemann tensor associated with the background geometry, and \({T}_{\mu\nu}\) corresponds to the energy momentum tensor. This equation is of great significance in understanding the effects of the trace-reversed metric and the associated force-free electromagnetic waves on the Kerr background \citep{ChandraPapers-RNpaperf}.

To conduct this study, we consider the prevailing physical conditions that allow us to focus on terms up to the order of \(\mathcal{O}(\lambda^2/L^2)\). Let us clarify this in more detail: our objective is to delve into the equations governing the evolution of a system under a specific configuration. Specifically, we seek to understand the disturbances that arise when force-free electrodynamics is present in a background characterized by a Kerr metric; a mathematical derivation is provided in Appendix \ref{Appx1}.

To streamline our calculations and gain valuable insights, we introduce two parameters that capture the scales of perturbations (\(\lambda\)) and the variations in the background metric (\(L\)). This assumption proves to be immensely helpful as it simplifies our analysis and leads to intriguing findings. By considering the relative magnitudes of \(\lambda\) and \(L\), we can effectively discern the dominant effects and the key features of the system's evolution. As a consequence, we can simplify equation (\ref{wavepropagation01}) by neglecting the second term, resulting in a more manageable form
\begin{equation}\label{wavepropagation02}
\square \bar{h}_{\mu \nu} = 4 {T}_{\mu\nu}\,,
\end{equation}
here we suppose \(G=1.\) This equation provides us with valuable insights into the propagation of gravitational waves in the presence of matter. To further explore this, we can utilize the radiative wave form and establish the following relationship
\begin{equation}\label{Waveform02}
\begin{array}{l}
\bar h_{\mu \nu}  =16 \pi \int \frac{d^3 R}{4 \pi|r-\mathbf{R}|} T_{\mu \nu}
 =16 \pi \int d^3 R\left(\sum_{l, m}T_{\mu \nu} \frac{1}{2 l+1} \frac{R^l}{r^{l+1}} Y_l^m(\theta, \varphi) Y_l^{* m}(\theta, \phi)\right)\,,
\end{array}
\end{equation}
to calculate different components of \(\bar h_{\mu\nu}\) in Eq.(\ref{Waveform02}), the following relations for harmonic oscillators are useful
\begin{eqnarray}\label{HOs00}\nonumber
Y_0^0=\frac{1}{\sqrt{4 \pi}}, \quad Y_2^0=\sqrt{\frac{5}{4 \pi}}\left(\frac{3}{2} \cos ^2 \theta-\frac{1}{2}\right)\,,\\
\cos ^2 \theta=\frac{2}{3}\left(\sqrt{\frac{4 \pi}{5}} Y_2^0+\frac{1}{2}\right), \quad \sin ^2 \theta=\frac{2}{3}\left(1-\sqrt{\frac{4 \pi}{5}} Y_2^0\right)\,.
\end{eqnarray}
In the magnetostatic limit, the Green’s function \(|\mathbf{r} - \mathbf{R}|^{-1}\) can be expanded in a multipole series. To leading order, we use the approximations \citep{Weinberg,Ryder}.
\begin{eqnarray}
|\mathbf{r}-\mathbf{R}|^{-1}=\left(r^2-2 \mathbf{r} \cdot \mathbf{R}+R^2\right)^{-1 / 2} \approx \frac{1}{r}\left(1-\frac{\mathbf{r} \cdot \mathbf{R}}{r^2}\right)+\cdots\,,\\
|\mathbf{r}-\mathbf{R}|=\left(r^2-2 \mathbf{r} \cdot \mathbf{R}+R^2\right)^{1 / 2} \approx r\left(1-\frac{\mathbf{r} \cdot \mathbf{R}}{r^2}\right)+\cdots\,.
\end{eqnarray}
These approximations enable us to simplify the mathematical expressions to obtain the modifications to the background metric satisfactorily.

\section{FFMF induced perturbations}\label{Perturbationshmunu}

With the help of the extracted energy momentum tensor, we are now in a position to obtain both the diagonal and off-diagonal components of \(\bar h_{\mu\nu}\), as appearing in Eq.(\ref{Waveform02}). Therefore first by utilizing the Eqs.(\ref{T44-Kerr}) to (\ref{T33-Kerr}) we will obtain the diagonal components of \(\bar h_{\mu\nu}\).

\begin{equation}\label{hbar44}
\begin{array}{l}
\bar h_{44}\simeq\frac{{32\pi M}}{{15{k^2}r}}\Big(\left( {9{a^2}{k^2} - 5{k^2}} \right)\ln \left( {kr} \right) - \left( {{a^2}{k^4} + 3{a^2}{k^2}} \right)\ln \left( {kr} \right)
- \left( {5 + \left( { - 9{a^2} + 5} \right){k^2}} \right)\ln \left( 2 \right)\\
 \hspace{1cm}- \left( { - 10 + 3\left( {4 - 3\gamma } \right){a^2} + 5\gamma } \right){k^2} - 5\gamma  + 10\Big)
 - \frac{{32}}{{15}}\pi {k^2}M\left( {{a^2} - \frac{5}{6}} \right)r - \frac{{16}}{{75}}\pi {k^4}M\left( {{a^2} - 5} \right){r^3}\,,
\end{array}
\end{equation}

\begin{equation}\label{hbar11}
\begin{array}{l}
\bar h_{11}\simeq
\frac{{{\rm{224}}\pi M}}{{15{k^3}r}}\Bigg[ - \frac{{10}}{7} + \left( {\left( {{a^2} - \frac{{20{M^2}}}{7} + \frac{5}{7}} \right){k^2}} \right)\ln \left( {kr} \right) + \left( {\frac{{\left( { - 9{a^2} + 20{M^2}} \right){k^4}}}{7} + \left( {{a^2} - \frac{{20{M^2}}}{7}} \right){k^2}} \right)\ln \left( {kr} \right)\\
\hspace{2cm} + \left( {\frac{5}{7} + \left( {{a^2} - \frac{{20{M^2}}}{7} + \frac{5}{7}} \right){k^2}} \right)\ln \left( 2 \right) + \gamma \left( {{a^2} - \frac{{20{M^2}}}{7} + \frac{5}{7}} \right){k^2} + \frac{{5\gamma }}{7}\Bigg] \\
 \hspace{2cm}- \frac{{16}}{{15}}\pi {k^2}M\left( {14{a^2} - 40{M^2} + 5} \right)r + \frac{{64}}{{135}}\left( {{a^2} - 10{M^2} + \frac{5}{4}} \right)\pi {k^2}{r^2}\,,
\end{array}
\end{equation}

\begin{equation}\label{hbar22}
\begin{array}{l}
\bar h_{22}\simeq
\frac{{4\pi \sin \left( {2kr} \right)\left( {\pi {a^2}{k^4} - 7{a^2}{k^4} - 15\,{a^2}{k^2} + 50{k^2} + 35} \right)}}{{15{k^5}r}} + \frac{{8\pi \sin \left( {2kr} \right)\left( {{a^2}{k^4} + 3{a^2}{k^2} - 40{k^2} - 35} \right)r}}{{15{k^3}}}\\
\\
\hspace{1cm}-\frac{{16\pi \left( {3{a^2}{k^2} + 60{k^2}{{\cos }^2}\left( {kr} \right) - 25{k^2} + 60{{\cos }^2}\left( {kr} \right) - 25} \right)}}{{45{k^2}}}{r^2}\,,
\end{array}
\end{equation}

\begin{equation}\label{hbar33}
\begin{array}{l}
\bar h_{33}\simeq- \frac{{32\pi \sin \left( {2kr} \right)\left( { - \frac{7}{4} - \frac{{{a^2}{k^4}}}{2} + \left( {\frac{{{a^2}}}{4} + 4} \right){k^2}} \right)}}{{15{k^5}r}} - \frac{{32\pi \sin \left( {2kr} \right)\left( {{a^2}{k^6} + \left( { - \frac{{{a^2}}}{2} - 8} \right){k^4} + \frac{{7{k^2}}}{2}} \right)}}{{15{k^5}}}r\\
\\
\hspace{1cm} + \frac{{32\pi \left( { - 40\left( {{k^4} - \frac{1}{2}{k^2}} \right)k\cos {{\left( {kR} \right)}^2} + \left( {\left( {\frac{{5{a^2}}}{3} + \frac{{50}}{3}} \right){k^4} - \frac{{25{k^2}}}{3}} \right)k} \right)}}{{75{k^5}}}{r^2}\,.
\end{array}
\end{equation}
Here \(\gamma\) is the renowned Euler-Mascheroni constant. Some interesting cases can be checked. For instance if we consider \(a=0\), then the corrections reduce to the Schwarzschild metric in the presence of an FFMF, which shows good agreement with previous results \citep{Sheikhahmadi}. If we switch off the FFMF by setting \(k=0\), then the problem immediately reduces to the Kerr solution. Now we can calculate the off-diagonal components.
\begin{equation}\label{hbar43}
\begin{array}{l}
\bar h_{34}=\bar h_{43}\simeq \frac{{32\pi Ma\left( {\left( { - 1 + \frac{9}{7}{a^2}{k^2} - 2{k^2}} \right)\left( {\gamma  + \ln \left( 2 \right) + \ln \left( {kr} \right)} \right) + 2 - \frac{{4{k^2}\left( {3{a^2} - 7} \right)}}{7} + \left( { - \frac{2}{7}{a^2}{k^4} - \frac{3}{7}{a^2}{k^2} + 1} \right)\ln \left( {kr} \right)} \right)}}{{15{k^2}r}}\\
\hspace{2cm} - \frac{{32}}{{105}}a\left( {{a^2} - \frac{7}{3}} \right)\pi {k^2}Mr - \frac{{16}}{{525}}\pi Ma{\mkern 1mu} {k^4}\left( {{a^2} - 14} \right){r^3}\,,
\end{array}
\end{equation}

\begin{equation}\label{hbar13}
\begin{array}{l}
\bar h_{13}=\bar h_{31}\simeq \frac{{4\pi \left( {15k - 30\cos {{\left( {kr} \right)}^2}k - 15\sin \left( {2kr} \right)M{k^2}} \right)}}{{75{k^4}r}} \\
\hspace{2cm}+ \frac{{4\pi \left( { - 10{k^3} + 20{k^3}\cos {{\left( {kr} \right)}^2} + 10\sin \left( {2kr} \right)M{\mkern 1mu} {k^4}} \right)}}{{75{k^4}}}r + \frac{{4\pi \left( {\frac{{5{k^5}M}}{4} + 10{k^4}\sin \left( {2kr} \right)} \right)}}{{75{k^4}}}{r^2}\,,
\end{array}
\end{equation}

\begin{equation}\label{hbar12}
\begin{array}{l}
\bar h_{12}=\bar h_{21}\simeq
\frac{{{\pi ^2}\left( {\left( {\left( { - {a^2}M + \frac{5}{2}{m^3}} \right){k^2} - \frac{{11M}}{2}} \right){k^2}{{\cos }^2}\left( {kr} \right) - 2\left( { - \frac{9}{4} + \frac{{9{k^2}{M^2}}}{8}} \right)\sin \left( {2kr} \right)k - \frac{{\left( {\left( { - {a^2}M + \frac{5}{2}{M^3}} \right){k^2} - \frac{{11M}}{2}} \right){k^2}}}{2}} \right)}}{{4{k^4}r}}\\
\\
\hspace{2cm} + \frac{{\left( {M{\mkern 1mu} k{{\cos }^2}\left( {kr} \right) - 2\sin \left( {2kr} \right) - \frac{{M{\mkern 1mu} k}}{2}} \right){\pi ^2}}}{{4k}}r + \frac{{\left( {{{\cos }^2}\left( {kr} \right) - \frac{1}{2}} \right){\pi ^2}}}{4}{r^2}\,,
\end{array}
\end{equation}

\begin{equation}\label{hbar23}
\begin{array}{l}
\bar h_{23}=\bar h_{23}\simeq
\frac{{{\pi ^2}\left( {\left( {\left( { - {a^2}M + \frac{5}{2}{M^3}} \right){k^2} - \frac{{11M}}{2}} \right){k^2}{{\cos }^2}\left( {kr} \right) - 2\sin \left( {2kr} \right)\left( { - \frac{9}{4} + \frac{{9{k^2}{M^2}}}{8}} \right)k - \frac{{\pi \left( {\left( { - {a^2}M + \frac{5}{2}{M^3}} \right){k^2} - \frac{{11M}}{2}} \right){k^2}}}{2}} \right)}}{{4{k^4}r}}\\
\\
\hspace{2cm} + \frac{{{\pi ^2}\left( {Mk\cos {^2}\left( {kr} \right) - 2\sin \left( {2kr} \right) - \frac{{Mk}}{2}} \right)}}{{4k}}r + \frac{{{\pi ^2}\left( {{{\cos }^2}\left( {kr} \right) - \frac{1}{2}} \right)}}{4}{r^2}\,.
\end{array}
\end{equation}
When considering the background Kerr metric, denoted as \(g_{\mu \nu}^{(0)}\), and examining the diagonal expressions for \(\bar h_{\mu\nu}\) (given by Eqs. (\ref{hbar44}) to (\ref{hbar33})), an interesting result emerges: the trace of \(\bar h_{\mu\nu}\) is found to be zero, i.e., \(\bar h = 0\). This implies that for the perturbed metric, we have \(h_{\mu\nu} = \bar h_{\mu\nu}\).
Upon closer analysis of Eqs. (\ref{hbar43}) to (\ref{hbar33}), we observe that in regions far from the black hole, and at the linearized order of the theory, the Kerr parameter does not appear explicitly. As the strength of the force-free magnetofluid diminishes, the associated corrections become negligible, approaching these \(C_{\mu\nu}\) coefficients. Finally in the limit \(k\rightarrow 0\), the solution will turn into the pure Kerr solution.
In the next sections we investigate some important physical properties of the explained configuration to see the role of the FFMF on the behaviour of a test particle, for instance, around this modified Kerr metric.

\section{Particle motion in the FFMF corrected Kerr metric}
\label{ParticleMotion}

In the present section we consider the motion of particles in the modified Kerr metric given by Eq.~(\ref{1-0}), where the perturbations appear explicitly in Eqs.~\eqref{hbar44} to \eqref{hbar23}. We obtain, and discuss in detail, the horizon radii, the effective potential, and the positions of the innermost stable circular orbits, see \cite{Vincent:2011wz,Virbhadra:1999nm,Yang:2018wye,Yang:2018wye,Sheikhahmadi:2023jpb}.

\subsection{Motion in arbitrary stationary and axially symmetric geometries}

Consider an arbitrary stationary and axially symmetric geometry, with the metric given in general form by \cite{Sheikhahmadi:2020snl,Soroushfar:2021mis}
\begin{equation}
ds^{2}=g_{tt}dt^{2}+g_{t\phi }dtd\phi +g_{rr}dr^{2}+g_{\theta \theta
}d\theta ^{2}+g_{\phi \phi }d\phi ^{2}.  \label{ds2rcoappr}
\end{equation}
We adopt the equatorial approximation by imposing \(|\theta -\pi /2|\ll 1\). Then all metric components \(g_{tt}\), \(g_{t\phi }\), \(g_{rr}\), \(g_{\theta \theta }\) and \(g_{\phi \phi }\) are functions of \(r\) only. For the metric (\ref{ds2rcoappr}) the geodesic equations are
\begin{equation}
\left( \frac{dt}{d\tau }\right) ^{2}=\frac{\widetilde{E}g_{\phi \phi }+%
\widetilde{L}g_{t\phi }}{g_{t\phi }^{2}-g_{tt}g_{\phi \phi }},
\end{equation}
\begin{equation}
\left( \frac{d\phi }{d\tau }\right) ^{2}=-\frac{\widetilde{E}g_{t\phi }+%
\widetilde{L}g_{tt}}{g_{t\phi }^{2}-g_{tt}g_{\phi \phi }},  \label{eq:phidot}
\end{equation}
and
\begin{equation}
g_{rr}\left( \frac{dr}{d\tau }\right) ^{2}=V(r),  \label{eq:rdot}
\end{equation}
where \(\tau\) is an affine parameter and the potential \(V(r)\) is
\begin{equation}\label{potentialKerr}
V(r)\equiv -1+\frac{\widetilde{E}^{2}g_{\phi \phi }+2\widetilde{E}\widetilde{%
L}g_{t\phi }+\widetilde{L}^{2}g_{tt}}{g_{t\phi }^{2}-g_{tt}g_{\phi \phi }}.
\end{equation}

In the equatorial plane, for circular orbits the conditions
\begin{equation}
V(r)=0,\qquad V_{,r}(r)=0
\end{equation}
must hold simultaneously. From these, the specific energy \(\widetilde{E}\), specific angular momentum \(\widetilde{L}\), and angular velocity \(\Omega\) of particles on circular orbits are
\begin{eqnarray}\label{Conserved-entities}
\widetilde{E} &=&-\frac{g_{tt}+g_{t\phi }\Omega }{\sqrt{-g_{tt}-2g_{t\phi
}\Omega -g_{\phi \phi }\Omega ^{2}}},  \label{tildeE} \\
\widetilde{L} &=&\frac{g_{t\phi }+g_{\phi \phi }\Omega }{\sqrt{%
-g_{tt}-2g_{t\phi }\Omega -g_{\phi \phi }\Omega ^{2}}},  \label{tildeL} \\
\Omega  &=&\frac{d\phi }{dt}=\frac{-g_{t\phi ,r}+\sqrt{(g_{t\phi
,r})^{2}-g_{tt,r}g_{\phi \phi ,r}}}{g_{\phi \phi ,r}}. \label{OmegaGen}
\end{eqnarray}
The ISCOs are determined by
\begin{equation}
V_{,rr}(r)=0.
\end{equation}

To simplify the formalism we write the effective potential as
\begin{equation}
V(r)\equiv -1+\frac{\tilde{f}}{\tilde{g}},
\end{equation}
with
\begin{equation}
\tilde{f}\equiv \widetilde{E}^{2}g_{\phi \phi }+2\widetilde{E}\widetilde{L}g_{t\phi
}+\widetilde{L}^{2}g_{tt},
\end{equation}
\begin{equation}
\tilde{g}\equiv g_{t\phi }^{2}-g_{tt}g_{\phi \phi }.
\end{equation}
Then \(V(r)\geq 0\) gives \(\tilde{f}=\tilde{g}\). The condition \(V_{,r}(r)=0\) yields
\begin{equation}
\tilde{f}_{,r}\tilde{g}-\tilde{f}\tilde{g}_{,r}=0,
\end{equation}
and \(V_{,rr}(r)=0\) leads to
\begin{equation}
\tilde{f}_{,rr}-\tilde{g}_{,rr}=0,
\end{equation}
where we used \(V=0\) and \(V_{,r}=0\). Thus, if \(\tilde{g}\neq 0\),
\begin{equation}
\widetilde{E}^{2}g_{\phi \phi ,rr}+2\widetilde{E}\widetilde{L}g_{t\phi ,rr}+%
\widetilde{L}^{2}g_{tt,rr}-(g_{t\phi }^{2}-g_{tt}g_{\phi \phi })_{,rr}=0. \label{stable}
\end{equation}
Once \(g_{tt}\), \(g_{t\phi}\) and \(g_{\phi \phi}\) are known, substituting Eqs.~(\ref{tildeE})-(\ref{tildeL}) into Eq.~(\ref{stable}) gives an algebraic equation for \(r\) whose solutions are the marginally stable orbits.

\subsection{Motion around the modified Kerr black hole}

\begin{figure}[ht]
\centering
\subfigure{\includegraphics[scale=0.33]{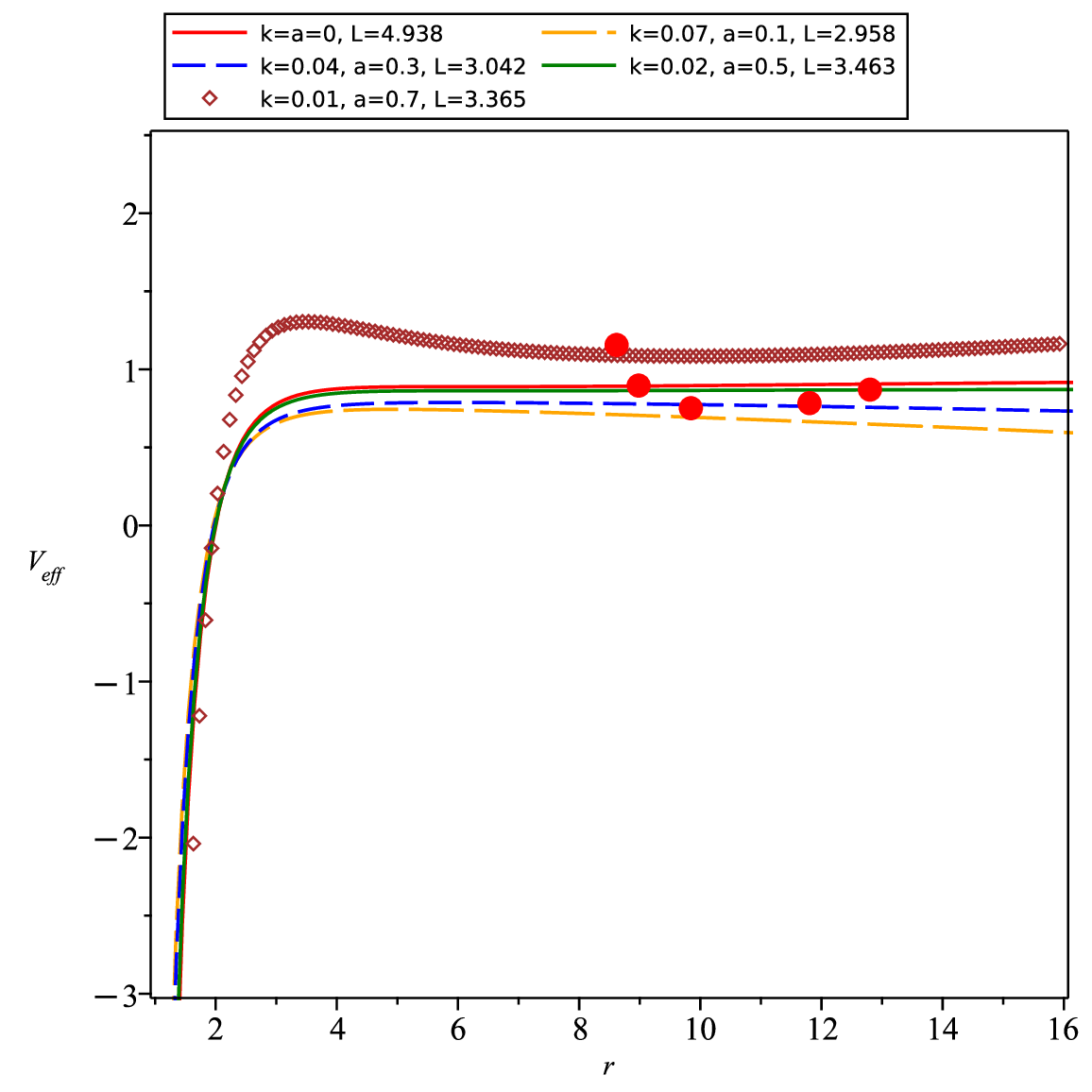}}
\subfigure{\includegraphics[scale=0.33]{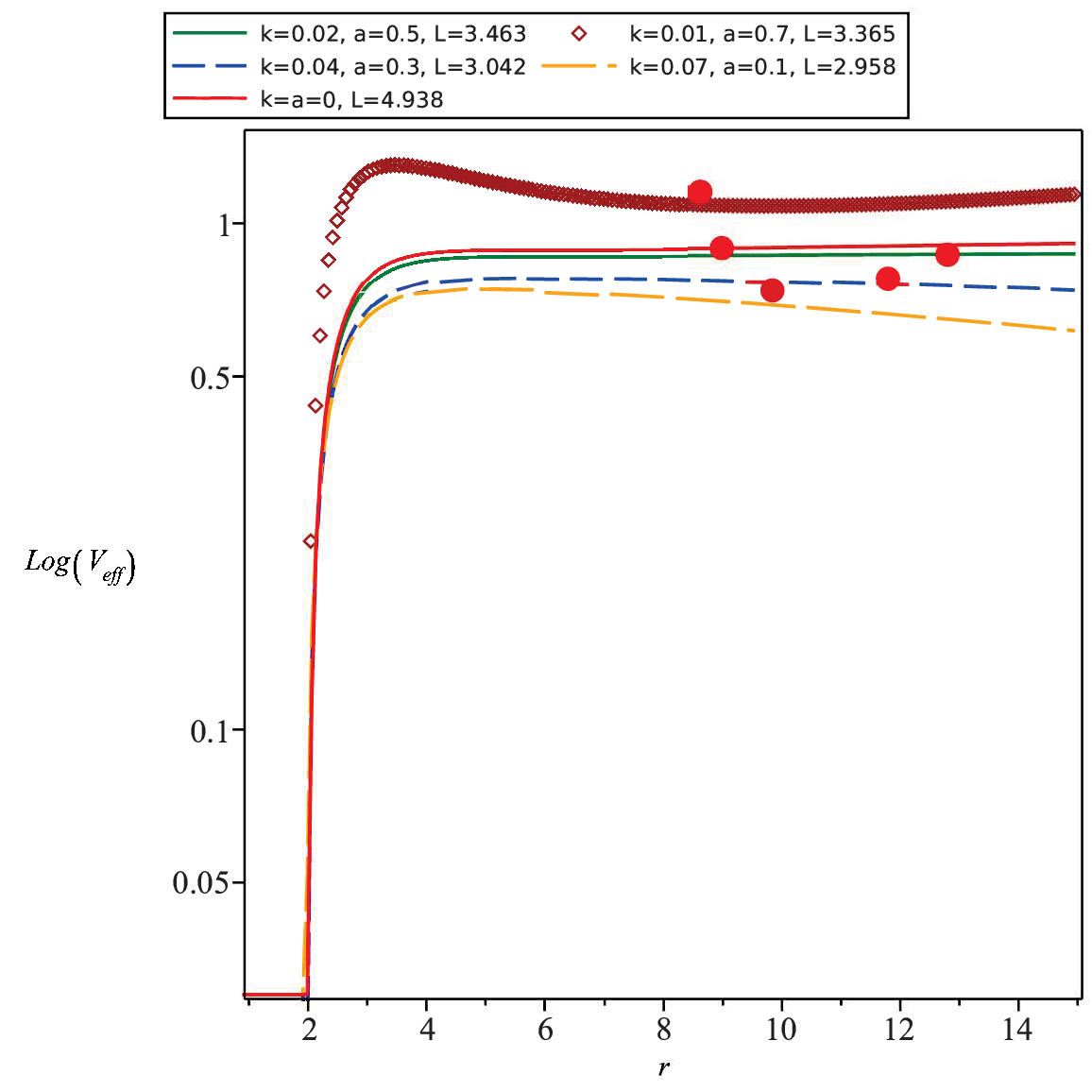}}
\caption{Behaviour of the effective potential versus \(r\) for different values of the parameters \(k\) and \(a\). The red bullets $(\rc{\bullet})$ indicate the location of the innermost stable circular orbits (ISCO).}
\label{fig:Veff0aaa}
\end{figure}

Now we investigate particle motion around the modified Kerr black hole. Using the metric (\ref{1-0}) we obtain \(\Omega\), \(\widetilde{E}\) and \(\widetilde{L}\) from Eq.~(\ref{Conserved-entities}). From these and Eq.~(\ref{potentialKerr}) for the effective potential we plot Fig.~\ref{fig:Veff0aaa} (see also \citep{Soroushfar:2021mis,Sheikhahmadi:2023jpb}).

\subsection{The case of the static spherically symmetric geometry}

As a simple example, and to simplify the comparison, consider a static spherically symmetric geometry with line element
\begin{equation}  \label{line}
ds^{2}=e^{\nu (r)}dt^{2}-e^{\lambda (r)}dr^{2}-r^{2}d\Omega ^{2},
\end{equation}
where \(d\Omega ^{2}=d\theta ^{2}+\sin ^{2}\theta d\varphi ^{2}\). The metric components are
\begin{equation}
g_{\mu \nu }=\mathrm{diag}(e^{\nu (r)},-e^{\lambda (r)},-r^{2},-r^{2}\sin
^{2}\theta ).
\end{equation}
In this case, for equatorial motion the geodesic equations become
\begin{eqnarray}
e^{2\nu }\left( \frac{dt}{d\tau }\right) ^{2} &=&\widetilde{E}^{2}, \\
e^{(\nu +\lambda )}\left( \frac{dr}{d\tau }\right) ^{2} &+&V_{\text{eff}}(r)=%
\widetilde{E}^{2}, \\
r^{4}\left( \frac{d\phi }{d\tau }\right) ^{2} &=&\widetilde{L}^{2}.
\end{eqnarray}
The effective potential is
\begin{equation}
V(r)\equiv e^{\nu }\left( 1+\frac{\widetilde{L}^{2}}{r^{2}}\right). \label{V2}
\end{equation}

The conditions for the innermost stable circular orbit \(V(r)=0\) and \(V_{,r}(r)=0\) give
\begin{eqnarray}
\Omega  &=&\sqrt{\frac{\nu _{,r}e^{\nu }}{2r}}=\sqrt{\frac{1}{2r}\frac{d}{dr}%
e^{\nu }}=\frac{1}{r_{g}}\sqrt{\frac{1}{2\eta }\frac{d}{d\eta }e^{\nu
(\eta )}},  \label{OmegaStatic} \\
\widetilde{E} &=&\frac{e^{\nu }}{\sqrt{e^{\nu }-r^{2}\Omega ^{2}}}=\frac{%
e^{\nu (\eta )}}{\sqrt{e^{\nu (\eta )}-\eta ^{2}\Omega ^{2}(\eta )}}, \label{Estat} \\
\widetilde{L} &=&\frac{r^{2}\Omega }{\sqrt{e^{\nu }-r^{2}\Omega ^{2}}}=r_{g}%
\frac{\eta ^{2}\Omega (\eta )}{\sqrt{e^{\nu (\eta )}-\eta ^{2}\Omega
^{2}(\eta )}},  \label{Lstat}
\end{eqnarray}
where \(\eta=r/r_g\). In dimensionless form,
\begin{equation}
V(\eta )\equiv e^{\nu (\eta )}\left( 1+\frac{\widetilde{L}^{2}(\eta )}{\eta
^{2}}\right).
\end{equation}
The marginally stable orbit radius \(r_{ms}\) follows from \(V_{,rr}(r)=0\).

\section{Thermal properties of accretion disk}
\label{Observ00}

\subsection{Flux and temperature of thin accretion disks}

Accreting black holes form accretion disks that are important sources of shadow information~\citep{Meng:2023uws,Virbhadra:1999nm}. We simulate photons emitted from the disk by tracing backward from the image plane to the disk plane. The radiation flux corresponds to the luminosity on the image plane. We use the Page-Thorne thin disk model for the flux \citep{Page:1974he}.
We first consider a general axially symmetric geometry as in Eq.\eqref{ds2rcoappr}, then we study the flux, temperature and efficiency parameter of the accretion disk around the FFMF modified Kerr metric.
The observed specific intensity at frequency \(\nu _{\mathrm{obs}}\) at point \((X,Y)\) is obtained by integrating the specific emissivity along the photon path~\citep{Pugliese:2025okq,Bambi:2013nla}
\begin{equation}
I_{\mathrm{obs}}(\nu _{\mathrm{obs}},X,Y)=\int_{\gamma }g^{3}j(\nu _{\mathrm{%
e}})dl_{\mathrm{prop}}, \label{eq-I}
\end{equation}
where \(g=\nu _{\mathrm{obs}}/\nu _{\mathrm{e}}\) is the redshift factor, \(\nu_{\mathrm{e}}\) the emitter rest-frame frequency, \(j(\nu _{\mathrm{e}})\) the emissivity per unit volume, and \(dl_{\mathrm{prop}}\) the proper length. The redshift factor is
\begin{equation}
g=\frac{k_{\alpha }u_{\mathrm{obs}}^{\alpha }}{k_{\beta }u_{\mathrm{e}}^{\beta }}=\frac{\sqrt{-g_{\phi \phi} \Omega^2-2 g_{t \phi} \Omega-g_{t t}}}{1+\frac{k_\phi}{k_t} \Omega},
\end{equation}
with \(k^{\mu}\) the photon 4-momentum, \(u_{\mathrm{obs}}^{\mu}=(1,0,0,0)\), and \(u_{\mathrm{e}}^{\mu}\) the emitter 4-velocity. For free-falling gas in a static spherically symmetric spacetime \cite{Pugliese:2025okq},
\begin{equation}
u_{\mathrm{e}}^{t}=\dot{t}=\frac{1}{\sqrt{-g_{\phi \phi} \Omega^2-2 g_{t \phi} \Omega-g_{t t}}},\quad u_{\mathrm{e}}^{r}=-\sqrt{\frac{%
V(r)}{g_{rr}}},\quad u_{\mathrm{e}}^{\theta }=u_{\mathrm{e}}^{\phi }=0.
\end{equation}
For monochromatic emission with rest frequency \(\nu_\star\) and a \(1/r^2\) profile,
\begin{equation}
j(\nu _{\mathrm{e}})\propto \frac{\delta (\nu _{\mathrm{e}}-\nu _{\star })}{%
r^{2}}.
\end{equation}
Also,
\begin{equation}
dl_{\mathrm{prop}}=k_{\alpha }u_{\mathrm{e}}^{\alpha }d\lambda=\frac{k_{t}}{g|k_{r}|}dr.
\end{equation}
Integrating Eq.~(\ref{eq-I}) over all frequencies gives the observed photon flux~\citep{Bambi:2013nla,Novikov:1973kta}
\begin{equation}
F_{\mathrm{obs}}(X,Y)\propto \int_{\gamma }\frac{g^{3}k_{t}dr}{r^{2}|k_{r}|}.
\end{equation}

\begin{figure}[ht]
\centering
\subfigure{\includegraphics[scale=0.63]{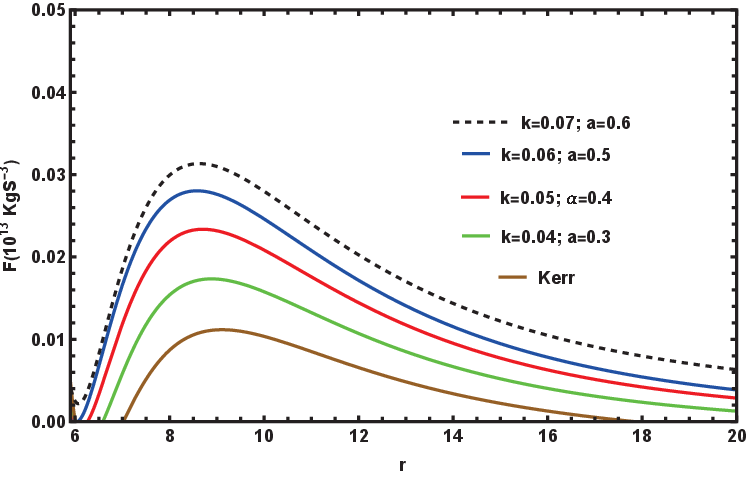}}
\subfigure{\includegraphics[scale=0.39]{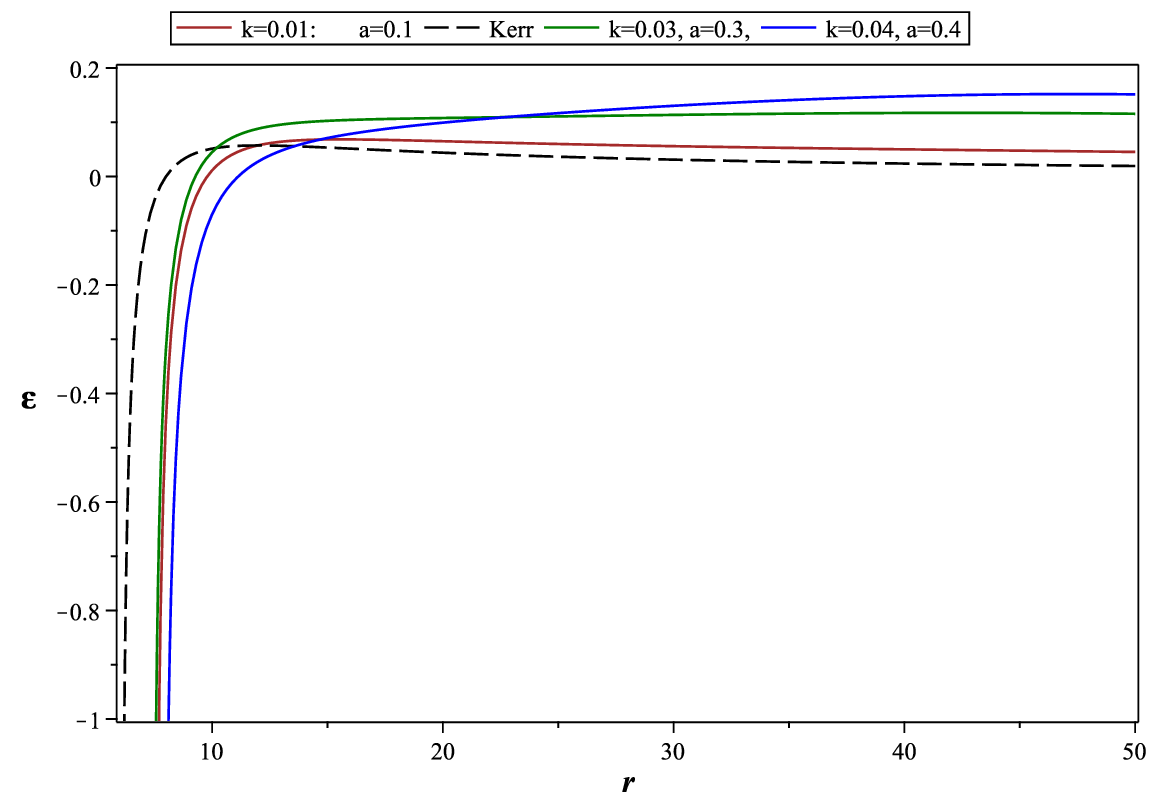}}
\caption{Flux (left) and efficiency parameter (right) of the accretion disk. Here \(G=c=1\); all curves: \(M=15M_\odot\). Here $\dot{M}_0 = 0.1\,\dot{M}_{\text{Edd}}$, with $\dot{M}_{\text{Edd}} = 1.4\times10^{18}\,(M/M_\odot)\,\text{g/s}$. For $M=15\,M_\odot$, this gives $\dot{M}_0 \approx 2.1\times10^{18}\,\text{g/s}$.Please note the subscript $Edd$ refers Eddington. }
\label{fig:flux00efficiency}
\end{figure}

For a geometrically thin accretion disk on the equatorial plane, the time-averaged radiant energy flux is \citep{Page:1974he}
\begin{equation}
F({r})=\dfrac{-\dot{M}_{0}}{4\pi \sqrt{-g}}\dfrac{\Omega _{,{r}}}{(\tilde{E}%
-\Omega {\tilde{L}})^{2}}\int_{r_{\text{isco}}}^{r}(\tilde{E}-\Omega {%
\tilde{L}}){\tilde{L}}_{,{r}}d{r}, \label{F}
\end{equation}
where \(\dot{M}_{0}\) is the time-averaged accretion rate and \(r_{\text{isco}}\) the ISCO radius (see Fig.~\ref{fig:flux00efficiency}). Assuming thermal equilibrium, from the Stefan–Boltzmann law \citep{Karimov:2018whx,Liu:2019mls},
\begin{equation}\label{Sigma}
 F(r)=\sigma T^{4}(r),
\end{equation}
with \(\sigma\) the Stefan–Boltzmann constant, we obtain the temperature \(T(r)\) (Fig.~\ref{fig:flux00efficiency}).

The efficiency parameter \(\mathcal{E}\), measuring the fraction of accreted mass converted to radiation, is \cite{Karimov:2018whx,Liu:2019mls}
\begin{equation}\label{Energy00}
\mathcal{E}=1-\tilde{E}(r_{\text{isco}}),
\end{equation}
where \(\tilde{E}\) is given by Eq.~(\ref{tildeE}). Fig.~\ref{fig:flux00efficiency} illustrates the efficiency for the modified Kerr geometry.

\begin{figure}[ht]
\centering
\includegraphics[scale=0.49]{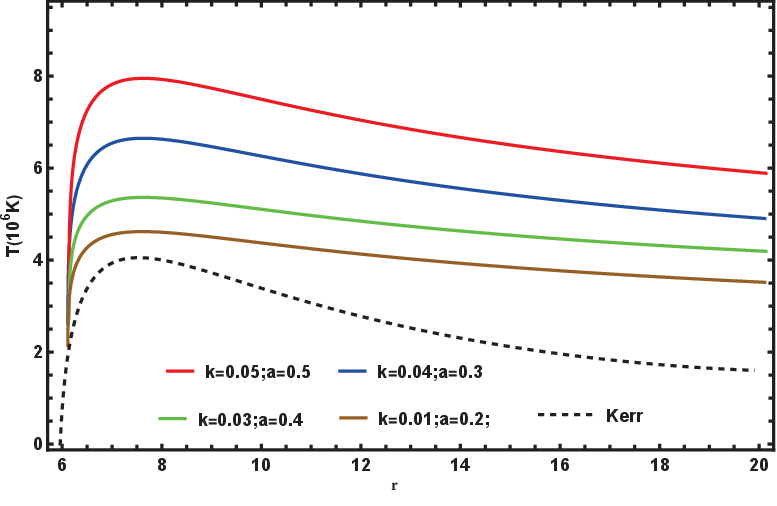}
\includegraphics[scale=0.44]{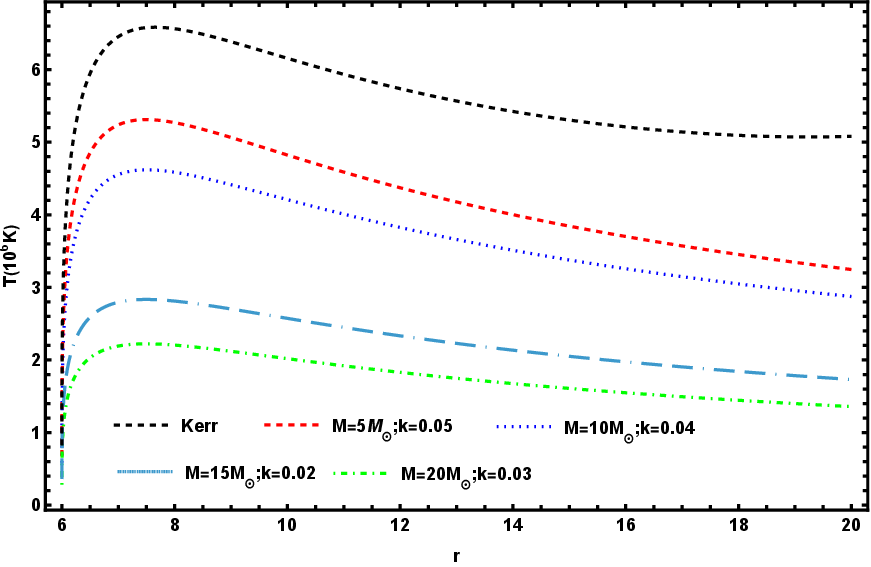}
\caption{Temperature is shown to see the effect of mass \(M\) and electromagnetic parameter \(k\). In the left panel we consider all black holes with the same mass, \(M=10 M_\odot\). In the right panel the mass varies but the rotation parameter is fixed at \(a=0.7\).}
\label{Pic:Efficiency}
\end{figure}

The redshift factor must be accounted for in the observed intensity. To obtain the shadow of the modified Kerr black hole, one traces photon trajectories from the observer back to the source. This is computationally intensive; we leave a detailed study for future work. For a detailed investigation of accretion disk and flux, see \citep{Harko:2009kj,Chen:2011wb,Yang:2018wye,Perez:2012bx,Karimov:2018whx,Liu:2019mls,Soroushfar:2021mis,Sheikhahmadi:2023jpb}.

\newpage
\section{Conclusion}\label{Final}

To elucidate the fundamental physics governing these systems, we conducted a detailed investigation addressing a central question: How do force-free magnetofluid sources modify the geometric properties of Kerr black holes? Specifically, we aimed to derive a metric that accurately describes magnetized accretion disks surrounding rotating black holes, providing a more realistic framework for astrophysical applications.

The primary objective was to characterize the interaction between force-free magnetofluids and black hole geometry, focusing on identifying metric modifications induced by magnetized accretion disks. Our approach yielded a refined metric that comprehensively describes the coupled system of a black hole and its surrounding magnetized plasma, offering significant insights into the complex physics of these astrophysical environments.

A key finding of this study is the inadequacy of perturbation theory based solely on the unperturbed background metric. Analysis of the perturbed Einstein field equations reveals that proper treatment requires additional metric corrections affecting both diagonal and off-diagonal components. These modifications provide a more complete description of the system's geometry.

The tetrad formalism proved essential in bridging flat and curved spacetime descriptions, enabling faithful reproduction of the electromagnetic field strength tensor in the Kerr background. This framework facilitated a deeper understanding of the interplay between warped spacetime and electromagnetic field configurations.

Our first-order perturbation calculations yielded metric corrections that significantly improve the system's description. These corrections exhibit distinct radial dependence: near the black hole (\(r \ll 1\)), they scale as \(1/r\), while in the far-field region (\(r \gg 1\)), they follow power-law behavior up to \(\mathcal{O}(r^3)\) for components \(h_{44}\) and \(h_{34}\). Notably, when both the Kerr parameter \(a\) and electromagnetic parameter \(k\) vanish, our solution reduces to the Schwarzschild metric, as expected. The asymptotic behavior bears resemblance to Kerr-de Sitter solutions, suggesting an effective cosmological constant-like contribution.

To see the effects of these modifications on the geodesic properties and thermal behaviour of the magnetised accretion disk around a Kerr metric, we investigated the ISCO and effective potential in detail, as shown in Fig.~\ref{fig:Veff0aaa}. The effective potential plots clearly indicate how the presence of the force-free magnetic field shifts the ISCO locations and modifies the stability of circular orbits. We then computed the energy flux, temperature distribution, and efficiency of the accretion disk. These results are presented in Figs.~\ref{fig:flux00efficiency} and \ref{Pic:Efficiency}, where we observe significant changes compared to the pure Kerr case: the flux and temperature profiles are altered, and the efficiency parameter is notably affected by both the spin \(a\) and the magnetic parameter \(k\).

Two promising research directions emerge from this work. First, extending this analysis to other types of black holes, such as charged black holes or higher-order ones immersed in force-free fields, could reveal novel phenomena arising from the interplay between black hole charge, accretion disk dynamics, and electromagnetic field configurations (see \citep{Sajadi:2023bwe,Sheikhahmadi:2020snl,Sheikhahmadi:2023jpb,Soroushfar:2021mis}). Such investigations would deepen our understanding of these complex systems.

Second, studying black hole shadows and other observational signatures, in addition to those we demonstrated here, in the presence of force-free fields offers substantial potential. The analysis of shadows, formed by gravitational light bending around black holes, has attracted considerable attention. Investigating these features in force-free environments could significantly enhance our understanding of black hole properties and their electromagnetic interactions \citep{Soroushfar:2021mis}, with direct implications for interpreting Event Horizon Telescope observations and similar data.

\appendix
\section{Explanation of the Short-Wavelength Approximation in Curved Spacetime}\label{Appx1}

In the study of wave propagation on curved backgrounds, a fundamental simplification occurs when the wavelength of the perturbation is much shorter than the characteristic curvature scale of the spacetime. This is the regime of geometric optics, in which waves propagate along null geodesics and their amplitudes satisfy transport laws that are decoupled from the detailed curvature of the background (see \citep{Isaacson1968a,Isaacson1968b,BrillHartle1964,BrillWheeler1957}). The purpose of this appendix is to provide a unified justification, grounded in the formalism of short-wavelength expansions, for the reduction of the fully coupled linearised wave equation
\begin{equation}\label{eq:fullwave}
\square \bar{h}_{\mu\nu} + 2 R^{(0)}_{\eta\mu\epsilon\nu} \bar{h}^{\eta\epsilon} = 4 T_{\mu\nu},
\qquad (\text{with } G=1)
\end{equation}
to the simpler form
\begin{equation}\label{eq:reducedwave}
\square \bar{h}_{\mu\nu} = 4 T_{\mu\nu},
\end{equation}
under the assumption \(\lambda \ll L\), where \(\lambda\) is the reduced wavelength of the perturbation and \(L\) is the smallest of the background curvature radius \(\mathcal{R}\) or the scale \(\mathcal{E}\) of amplitude variation.

The argument follows the classical two-length-scale expansion familiar from geometric optics \cite{ChandraPapers-RNpapereb}, and is illustrated elegantly by the comparison of exact electromagnetic and gravitational plane waves in a certain class of metrics. We shall first recall the relevant expansion scheme, then apply it to the wave equation, and finally connect the result to the illustrative example of the metric
\begin{equation}\label{eq:metric3540}
ds^2 = L^2(u)(dx^2+dy^2) - du\, dv, \qquad u = t-z,\; v = t+z,
\end{equation}
which appears in the study of plane waves in general relativity (see equation (35.40) of \cite{ChandraPapers-RNpapereb}).

\subsection*{The short-wavelength expansion}

Let \(\lambda\) denote the reduced wavelength of the perturbation (gravitational or electromagnetic) as measured in a local Lorentz frame, and let \(L\) be the minimum of the background curvature radius \(\mathcal{R}\) and the scale \(\mathcal{E}\) of variation of the wave amplitude. The geometric-optics condition is
\begin{equation}\label{eq:geometricopticscond}
\lambda \ll L .
\end{equation}
Introduce a formal small parameter \(\epsilon \sim \lambda/L\). Following the standard WKB-type ansatz \cite{ChandraPapers-RNpapereb}, we write the trace‑reversed metric perturbation as
\begin{equation}\label{eq:ansatz}
\bar{h}_{\mu\nu} = \Re\left\{\bigl( a_{\mu\nu} + \epsilon\, b_{\mu\nu} + \epsilon^2 c_{\mu\nu} + \dots \bigr) 
\, e^{i\theta/\epsilon} \right\},
\end{equation}
where \(\theta\) is a real phase function satisfying \(k_\mu \equiv \nabla_\mu\theta \neq 0\), and the amplitudes \(a_{\mu\nu}, b_{\mu\nu}, \dots\) are slowly varying complex tensors. The factor \(e^{i\theta/\epsilon}\) oscillates on the scale \(\lambda\), while the amplitudes vary on the scale \(L\).

\subsection*{Order‑by‑order analysis of the wave equation}

Insert the ansatz \eqref{eq:ansatz} into the full wave equation \eqref{eq:fullwave}. The covariant wave operator acts as
\begin{equation}\label{eq:waveopexpansion}
\square \bar{h}_{\mu\nu} = \Re\Bigl\{ \bigl[ -\epsilon^{-2} k^\alpha k_\alpha \, a_{\mu\nu}
+ \epsilon^{-1} (\text{transport terms}) + O(1) \bigr] e^{i\theta/\epsilon} \Bigr\}.
\end{equation}
The curvature‑coupling term is
\begin{equation}\label{eq:curvatureexpansion}
2 R^{(0)}_{\eta\mu\epsilon\nu} \bar{h}^{\eta\epsilon}
= \Re\Bigl\{ \bigl[ 2 R^{(0)}_{\eta\mu\epsilon\nu} a^{\eta\epsilon} + O(\epsilon) \bigr] e^{i\theta/\epsilon} \Bigr\},
\end{equation}
because the Riemann tensor of the background varies only on the scale \(L\) and therefore contains no explicit powers of \(\epsilon^{-1}\).

Collecting terms order by order in \(\epsilon\):

\paragraph{Order \(\epsilon^{-2}\):}
\begin{equation}\label{eq:eikonal}
k^\alpha k_\alpha \, a_{\mu\nu} = 0 \quad\Longrightarrow\quad k^\alpha k_\alpha = 0,
\end{equation}
the {eikonal equation}, stating that wavefronts are null surfaces.

\paragraph{Order \(\epsilon^{-1}\):}
Gives a transport equation for \(a_{\mu\nu}\) along the rays; it involves derivatives of the amplitude but not the curvature coupling.

\paragraph{Order \(\epsilon^{0}\) (i.e. \(O(1)\)):}
At this order the equation reads
\begin{equation}\label{eq:orderone}
\text{(terms from \(\square \bar{h}\) of order 1)} \;+\; 2 R^{(0)}_{\eta\mu\epsilon\nu} a^{\eta\epsilon}
= 4 T_{\mu\nu}.
\end{equation}
However, the crucial observation is that the curvature term \(2 R^{(0)}_{\eta\mu\epsilon\nu} a^{\eta\epsilon}\) is itself of order \(\epsilon^{2}\) {relative to the leading wave‑operator term} \(-\epsilon^{-2}k^\alpha k_\alpha a_{\mu\nu}\). Indeed, comparing \eqref{eq:waveopexpansion} and \eqref{eq:curvatureexpansion}:
\begin{equation}\label{eq:ratioscale}
\frac{|R^{(0)} a|}{|k^\alpha k_\alpha a|/\epsilon^{2}} \sim 
\frac{|R^{(0)}| \lambda^{2}}{|k|^2} \sim \frac{\lambda^{2}}{\mathcal{R}^{2}} \sim \epsilon^{2} \ll 1.
\end{equation}
Thus, to leading nontrivial order in the geometric‑optics expansion—i.e., keeping terms up to \(O(1)\) after factoring out the rapid phase—the curvature coupling term is negligible. What remains is simply
\begin{equation}\label{eq:finalreduced}
\square \bar{h}_{\mu\nu} = 4 T_{\mu\nu},
\end{equation}
provided the source \(T_{\mu\nu}\) itself does not contain terms enhanced by inverse powers of \(\epsilon\). This is the desired simplification, justifying the transition from \eqref{eq:fullwave} to \eqref{eq:reducedwave}.

\subsection*{Illustration with the plane‑wave metric}

The metric \eqref{eq:metric3540} offers a concrete realisation of this scale separation. Consider an electromagnetic potential \(A = \mathcal{A}(u)\, dx\) in this background. The corresponding field strength \(F_{ux} = \mathcal{A}'(u)\) describes a plane wave propagating in the \(z\)-direction. Its stress‑energy tensor has only one non‑vanishing component,
\begin{equation}\label{eq:Tuu}
T_{uu} = \frac{(\mathcal{A}')^{2}}{4\pi L^{2}} .
\end{equation}
Now impose the Einstein equations \(G_{\mu\nu}=8\pi T_{\mu\nu}\) with this source. For the metric \eqref{eq:metric3540} the relevant equation reduces to \cite{ChandraPapers-RNpapereb}
\begin{equation}\label{eq:Lequation}
L'' + 4\pi T_{uu} L = 0 .
\end{equation}
Remarkably, this is precisely the same equation that governs the “background factor” \(L(u)\) in the exact gravitational plane‑wave solution when the “ripple factor” \(\beta(u)\) is absent. In the short‑wavelength limit \(\lambda\to0\) while keeping the averaged energy density \(\langle T_{uu}\rangle\) fixed, the oscillatory part (\(\beta\)) becomes irrelevant and only the slowly varying background \(L(u)\) determines the net curvature felt by test particles. Consequently, the wave propagates as if the curvature term were absent from the wave equation, exactly the decoupling derived above.

This example underscores the physical meaning of the approximation: when the wavelength is sufficiently short, the wave does not resolve the detailed curvature of the background; instead it experiences only an effective, slowly varying geometry described by the background factor \(L(u)\). The curvature‑coupling term in the wave equation, being proportional to the Riemann tensor, is therefore suppressed by \((\lambda/\mathcal{R})^{2}\) relative to the leading derivatives, and can be dropped to leading order.

Therefore we have shown, using the systematic short‑wavelength expansion of geometric optics, that the linearised wave equation in a curved background simplifies to the covariant wave equation without explicit curvature coupling, \eqref{eq:finalreduced}, under the condition \(\lambda \ll L\). The argument is rigorous within the context of two‑length‑scale asymptotic analysis and is vividly exemplified by the behaviour of plane waves in the metric \eqref{eq:metric3540}. This reduction is essential for studying the propagation of force‑free electromagnetic waves on Kerr backgrounds, where the high‑frequency limit permits one to disregard the intricate coupling between the wave and the background Riemann tensor, thereby rendering the problem tractable while retaining the essential physics of wave propagation in strong gravitational fields.

\section{General metric and symmetries}\label{Appx2}

We consider a stationary, axisymmetric spacetime with two Killing vectors \(\partial_t\) and \(\partial_\phi\). In adapted coordinates \((t,r,\theta,\phi)\) the metric can be written in the form of Eq.~(\ref{ds2rcoappr}), where all metric functions are independent of \(t\) and \(\phi\). We restrict attention to the \textbf{equatorial plane} \(\theta = \pi/2\) and consider only \textbf{circular orbits} (\(\dot r = 0,\ \dot\theta = 0\)). Hence the metric functions become functions of \(r\) only:
\[
g_{tt}(r),\quad g_{t\phi}(r),\quad g_{rr}(r),\quad g_{\phi\phi}(r),\qquad g_{\theta\theta}(r) \text{ will not enter the dynamics.}
\]
\subsection*{Conserved quantities and radial equation}
For a test particle of unit mass, the four-velocity is \(u^\mu = (\dot t, 0, 0, \dot\phi)\) with \(\dot{} = d/d\tau\) (proper time). The two Killing vectors give two conserved specific quantities appearing in Eqs.\eqref{tildeE} and \eqref{tildeL} respectively. These are the specific energy and specific angular momentum, respectively.
\begin{align}
\tilde{E} &\equiv -g_{tt}\dot t - g_{t\phi}\dot\phi, \label{tildeE0} \\
\tilde{L} &\equiv  g_{t\phi}\dot t + g_{\phi\phi}\dot\phi. \label{tildeL0}
\end{align}
Solving (\ref{tildeE0}) and (\ref{tildeL0}) for \(\dot t\) and \(\dot\phi\) yields
\begin{align}
\dot t &= \frac{-g_{\phi\phi} \tilde{E} - g_{t\phi} \tilde{L}}{\Delta_g}, \label{eq:ttdot} \\
\dot\phi &= \frac{ g_{t\phi} \tilde{E} + g_{tt} \tilde{L}}{\Delta_g}, \label{eq:phidot0}
\end{align}
where the determinant of the \(t\)--\(\phi\) sector is
\[
\Delta_g \equiv -g_{tt}g_{\phi\phi} - g_{t\phi}^2 > 0.
\]
The normalization \(g_{\mu\nu}u^\mu u^\nu = -1\) gives
\[
g_{rr}\dot r^2 = -1 - \bigl( g_{tt}\dot t^2 + 2g_{t\phi}\dot t\dot\phi + g_{\phi\phi}\dot\phi^2 \bigr).
\]
Substituting Eqs.\eqref{eq:ttdot} and \eqref{eq:phidot0} and simplifying leads to the radial equation
\[
\dot r^2 = \frac{1}{g_{rr}}\left( \frac{g_{\phi\phi}\tilde{E}^2 + 2g_{t\phi}\tilde{E}\tilde{L} + g_{tt}\tilde{L}^2}{\Delta_g} - 1 \right). \label{eq:radeq0}
\]
Define the function
\[
V(r) \equiv \frac{g_{\phi\phi}\tilde{E}^2 + 2g_{t\phi}\tilde{E}\tilde{L} + g_{tt}\tilde{L}^2}{\Delta_g} - 1,
\]
so that \(\dot r^2 = V(r)/g_{rr}\). For circular orbits we require
\[
V(r) = 0 \quad\text{and}\quad V'(r) = 0. \label{eq:circ0}
\]
From \(V=0\) we obtain
\[
g_{\phi\phi}\tilde{E}^2 + 2g_{t\phi}\tilde{E}\tilde{L} + g_{tt}\tilde{L}^2 = \Delta_g. \label{eq:R00}
\]
Differentiating with respect to \(r\) and using \(V'=0\) gives
\[
g_{\phi\phi}' \tilde{E}^2 + 2g_{t\phi}' \tilde{E}\tilde{L} + g_{tt}' \tilde{L}^2 = \Delta_g'. \label{eq:Rprime0}
\]
These equations are two linear ones in the three quantities \(\tilde{E}^2\), \(\tilde{L}^2\) and \(\tilde{E}\tilde{L}\). Using the condition \(d V/dr = 0\) (or equivalently \(d\tilde{E}/dr = 0\) along the sequence of circular orbits) one can derive an expression for the angular velocity \(\Omega = \dot\phi/\dot t\):
\[
\Omega_{\pm}(r) = \frac{-g_{t\phi}' \pm \sqrt{(g_{t\phi}')^2 - g_{tt}' g_{\phi\phi}'}}{g_{\phi\phi}'}. \label{eq:Omega0}
\]
The \(+\) sign corresponds to prograde orbits (angular momentum aligned with the spin of the central object), the \(-\) sign to retrograde orbits. This formula holds provided \(g_{\phi\phi}'\neq 0\).
From the definitions of \(\tilde{E}\) and \(\tilde{L}\), one can also express them in terms of \(\Omega\) as Eqs. \eqref{tildeE} and \eqref{tildeL}.
\subsection*{Marginally bound orbit (MBO)}
A circular orbit is \textbf{marginally bound} if the specific energy equals the rest mass energy, i.e. \(\tilde{E}=1\) (in units where \(c=1\)). Using (\ref{tildeE}) we obtain
\[
\frac{-g_{tt} - g_{t\phi}\Omega}{\sqrt{-\bigl(g_{tt}+2g_{t\phi}\Omega+g_{\phi\phi}\Omega^2\bigr)}} = 1.
\]
Squaring and rearranging leads to
\[
(g_{tt}+g_{t\phi}\Omega)^2 = g_{tt}+2g_{t\phi}\Omega+g_{\phi\phi}\Omega^2.
\]
After simplification this becomes a quadratic in \(\Omega\):
\begin{equation}
\bigl(g_{t\phi}^2 - g_{\phi\phi}\bigr)\Omega^2 + 2g_{t\phi}(g_{tt}-1)\Omega + g_{tt}(g_{tt}-1) = 0. \label{eq:MBOeq0}
\end{equation}
The MBO radius \(r_{\text{MBO}}\) is obtained by solving (\ref{eq:MBOeq0}) together with the relation \(\Omega = \Omega_{\pm}(r)\) from (\ref{OmegaGen}). In practice one eliminates \(\Omega\) to obtain an algebraic equation in \(r\) alone.
\subsection*{Innermost stable circular orbit (ISCO)}
The ISCO is the smallest radius at which a stable circular orbit exists. Radial stability is determined by the second derivative of the effective potential. For a sequence of circular orbits, one can use the fact that \(\tilde{E}^2(r)\) (or \(V_{\text{eff}}\) along the sequence) has a minimum at the ISCO. Hence the condition is
\[
\frac{d\tilde{E}}{dr} = 0 \quad\Longleftrightarrow\quad \frac{d}{dr}\left( \frac{-g_{tt} - g_{t\phi}\Omega}{\sqrt{-\bigl(g_{tt}+2g_{t\phi}\Omega+g_{\phi\phi}\Omega^2\bigr)}} \right) = 0. \label{eq:ISCOcond0}
\]
An equivalent and often more practical approach starts from the three conditions for a marginally stable circular orbit:
\[
{V}(r) = 0,\qquad V'(r) = 0,\qquad V''(r) \geq 0.
\]
Treating \(\tilde{E}\) and \(\tilde{L}\) as functions of \(r\) determined by the first two conditions, the third gives a single algebraic equation for \(r\). After eliminating \(\tilde{E}\) and \(\tilde{L}\), one arrives at the compact formula
\begin{equation}
\bigl( g_{tt}' + 2g_{t\phi}'\Omega + g_{\phi\phi}'\Omega^2 \bigr)^2 = \frac12 \bigl( g_{tt} + 2g_{t\phi}\Omega + g_{\phi\phi}\Omega^2 \bigr) \bigl( g_{tt}'' + 2g_{t\phi}''\Omega + g_{\phi\phi}''\Omega^2 \bigr), \label{eq:ISCOfinal0}
\end{equation}
where \(\Omega = \Omega_{\pm}(r)\) is given by (\ref{OmegaGen}). Equation (\ref{eq:ISCOfinal0}) is the general condition for the ISCO.
These equations are generally algebraic and may require numerical solution for a given metric.
\subsection{Examples}
\subsubsection*{Schwarzschild metric}
For Schwarzschild (\(a=0\)): \(g_{tt}= -(1-2M/r)\), \(g_{t\phi}=0\), \(g_{\phi\phi}=r^2\). Then
\[
\Omega = \sqrt{\frac{M}{r^3}},\qquad \tilde{E} = \frac{1-2M/r}{\sqrt{1-3M/r}}.
\]
\begin{itemize}
\item \(\tilde{E}=1\) gives \(r=4M\) (MBO).
\item \(d\tilde{E}/dr=0\) gives \(r=6M\) (ISCO).
\end{itemize}
\subsubsection*{Kerr metric}
For Kerr (\(a = J/M\), \(0\le a \le M\)): \(g_{tt}= -\bigl(1-\frac{2Mr}{\rho^2}\bigr)\), \(g_{t\phi}= -\frac{2Mar}{\rho^2}\sin^2\theta\), \(g_{\phi\phi}= \bigl(r^2+a^2+\frac{2Ma^2r}{\rho^2}\sin^2\theta\bigr)\sin^2\theta\), restricted to \(\theta=\pi/2\) gives \(\rho^2=r^2\). The known results are recovered
\begin{align}
r_{\text{ISCO}} &= M\Bigl[3+Z_2 \mp \sqrt{(3-Z_1)(3+Z_1+2Z_2)}\Bigr],\\
r_{\text{MBO}} &= 2M \mp a + 2\sqrt{M(M\mp a)},
\end{align}
with the upper sign for prograde, lower for retrograde, and \(Z_1,Z_2\) defined as follows
\begin{eqnarray}
Z_1 &\equiv 1 + (1-a_*^2)^{1/3} \left[ (1+a_*)^{1/3} + (1-a_*)^{1/3} \right], \label{eq:Z1} \\
Z_2 &\equiv \sqrt{3a_*^2 + Z_1^2}, \label{eq:Z2} 
\end{eqnarray}
and \(a_* = a/M\) is the dimensionless spin parameter.

\section*{Acknowledgments}
H.Sh. wishes to thank the anonymous referee for insightful comments that greatly enhanced the clarity and presentation of this work. He also acknowledges Y. Sobouti for stimulating discussions during the initial drafting phase. Furthermore, he is grateful to H. Firouzjahi and A. Abebe for valuable conversations on black hole physics and for suggestions that improved the manuscript.

\section*{Declaration of competing interest}
The author declares that he has no known competing financial interests or personal relationships that could have appeared to influence the work reported in this paper.

\section*{CRediT authorship contribution statement}
Haidar Sheikhahmadi: Conceptualization, Investigation, Methodology, Visualization, Writing - original draft, review \& editing.

\section*{Data Availability Statement}
This manuscript has no associated data or the data will not be deposited. [Authors’ comment: The datasets generated during and/or analysed during the current study are available from the corresponding author on reasonable request.]

\end{document}